\documentclass[showpacs,amsmath,amssymb,aps,pra,
10pt,reprint,superscriptaddress,showkeys]{revtex4-1}
\usepackage{graphicx}
\usepackage{bm}
\usepackage[breaklinks=true,colorlinks=true,linkcolor=blue,urlcolor=blue,citecolor=blue]{hyperref}
\usepackage{graphics}
\usepackage{dcolumn}
\usepackage{amsmath,amssymb}
\usepackage{mathdots}
\usepackage{natbib}
\usepackage{soul,color}
\usepackage{epstopdf}
\usepackage[note-name]{notes2bib}

\begin{document}
\title{Spin-wave spectroscopy of individual ferromagnetic nanodisks}

\author{O. V.~Dobrovolskiy}
    \email{oleksandr.dobrovolskiy@univie.ac.at}
    \affiliation{Faculty of Physics, University of Vienna, Vienna, Austria}
\author{S. A.~Bunyaev}
    \affiliation{Institute of Physics for Advanced Materials, Nanotechnology and Photonics (IFIMUP)/Departamento de F\'isica e Astronomia,
    Universidade do Porto, Porto, Portugal}
\author{N.~R.~Vovk}
    \affiliation{Institute of Physics for Advanced Materials, Nanotechnology and Photonics (IFIMUP)/Departamento de F\'isica e Astronomia,
    Universidade do Porto, Porto, Portugal}
    \affiliation{Department of Physics, V. N. Karazin Kharkiv National University, Kharkiv, Ukraine}
\author{D.~Navas}
    \affiliation{Institute of Physics for Advanced Materials, Nanotechnology and Photonics (IFIMUP)/Departamento de F\'isica e Astronomia,
    Universidade do Porto, Porto, Portugal}
    \affiliation{Instituto de Ciencia de Materiales de Madrid, ICMM-CSIC, Madrid, Spain}
\author{P.~Gruszecki}
    \affiliation{Faculty of Physics, Adam Mickiewicz University in Pozna\'n, Pozna\'n, Poland}
    \affiliation{Institute of Molecular Physics, Polish Academy of Sciences, Pozna\'n, Poland}
\author{M.~Krawczyk}
    \affiliation{Faculty of Physics, Adam Mickiewicz University in Pozna\'n, Pozna\'n, Poland}
\author{R. Sachser}
    \affiliation{Institute of Physics, Goethe University, Frankfurt am Main, Germany}
\author{M.~Huth}
    \affiliation{Institute of Physics, Goethe University, Frankfurt am Main, Germany}
\author{A. V.~Chumak}
    \affiliation{Faculty of Physics, University of Vienna, Vienna, Austria}
\author{K.~Y.~Guslienko}
    \affiliation{Depto. Fisica de Materiales, Universidad del Pais Vasco, UPV/EHU, San Sebastian, Spain}
    \affiliation{IKERBASQUE, the Basque Foundation for Science, Bilbao, Spain}
\author{G.~N.~Kakazei}
    \affiliation{Institute of Physics for Advanced Materials, Nanotechnology and Photonics (IFIMUP)/Departamento de F\'isica e Astronomia,
    Universidade do Porto, Porto, Portugal}

\begin{abstract}
The increasing demand for ultrahigh data storage densities requires development of 3D magnetic nanostructures. In this regard, focused electron beam induced deposition (FEBID) is a technique of choice for direct-writing of various complex nano-architectures. However, intrinsic properties of nanomagnets are often poorly known and can hardly be assessed by local optical probe techniques. Here, we demonstrate spatially resolved spin-wave spectroscopy of individual circular magnetic elements with radii down to $100$\,nm. The key component of the setup is a microwave antenna whose $2\times4\,\mu$m$^2$ central part is placed over a movable substrate with well-separated CoFe-FEBID nanodisks. The circular symmetry of the disks gives rise to standing spin-wave resonances and allows for the deduction of the saturation magnetization and the exchange stiffness of the material using an analytical theory. The presented approach is especially valuable for the characterization of direct-write elements opening new horizons for 3D nanomagnetism and magnonics.
\end{abstract}

\maketitle

Patterned nanomagnets have traditionally been two-dimensional planar structures. Recent work, however, is expanding nanomagnetism into the third dimension \cite{Str15nac,Pac17nac,Win19jap}. This expansion is triggered by the development of advanced synthesis methods \cite{Wil18nar,Hut18mee} and the discovery of novel geometry- and topology-induced effects \cite{Str16jpd}. Among the various nanofabrication techniques, focused electron beam-induced deposition (FEBID) has unique advantages for 3D nanomagnetism \cite{Pac17nac,Win19jap, Sko20nal}, magnonics \cite{Gub19boo,Dob19nph} and fluxonics \cite{Dob17pcs,Por19acs} as the magnetic properties of direct-write structures can be varied by selecting a writing strategy and beam parameters \cite{Hut18mee}, as well as via post-growth irradiation of structures with ions \cite{Lar14apl,Dob19ami,Fla20prb} and electrons \cite{Dob15bjn}. In the context of superconductivity, hybrid ferromagnet/superconductor structures are used for the manipulation of magnetic flux quanta \cite{Dob11pcs,Dob19pra} and open access to investigations of odd-frequency spin-triplet superconductivity \cite{Kom14apl,Mel16prl}. In the domain of magnonics, FEBID allows for the realization of 3D magnonic crystals \cite{Kra14pcm,Chu17jpd,Beg18apl} and magnonic conduits with graded refractive index \cite{Dav15prb}. However, the magnetic parameters of individual FEBID nanoelements are often poorly known and can hardly be assessed by local optical probe techniques, such as Brillouin light scattering (BLS) \cite{Seb15fph} and magneto-optical Kerr microscopy \cite{Urs16aip}, having a spatial resolution limit of about $300\,$nm. For resolving the magnetization dynamics at GHz frequencies on the length scale of a few tens of nm more sophisticated techniques are required. For instance, near-field BLS allows for imaging the edge mode dynamics with a resolution of about $50$\,nm \cite{Jer10apl} and X-ray microscopy provides dynamic magnetization imaging with a resolution as fine as $30$\,nm \cite{Slu19nan,Gua20nac}. At the same time, these techniques are complex and not widely accessible. Therefore, the development of easy-to-use experimental techniques for the deduction of magnetic properties of individual nanoelements is of crucial importance.

Ferromagnetic resonance (FMR) is the method of choice for a quantitative analysis of saturation magnetization, magnetic anisotropy, dipolar interactions, relaxation of magnetization, and the structural quality and (in)homogeneity of magnetic materials \cite{Far98rpp,Hei91jap,Kal06jap,Kak06prb}. However, the sensitivity of FMR is often not high enough for the detection of signals from individual magnetic nanoelements. The sensitivity of the cavity-based FMR is estimated as $\simeq10^{10}$ spins and it can reach $10^{8}$--$10^{9}$ spins for its broadband strip-line counterpart \cite{Poo83boo}. In order to compensate for poor signal-to noise ratio (SNR), usually many magnetic elements are arranged in arrays \cite{Ali09prb,Zho15prb}. However, in case of ensemble measurements the signal analysis is often complicated by interactions between the individual elements and by the distribution in their sizes and properties. This is why a number of alternative FMR-based approaches have been introduced for studying magnetization dynamics in single nanomagnets, see e.g. Refs. \cite{Nem13prl,Saf16apl}. Among these approaches, FMR force microscopy (FMRFM) \cite{Lou09prl,Guo13prl} and microresonator-based FMR (MRFMR) \cite{Mol14sim,Sch17rsi} have a high sensitivity of $\simeq 10^6$ spins and allow for probing $\simeq 100$\,nm-large areas in thin-film samples \cite{Guo13prl,Sch17rsi}. However, separate resonators are needed for each individual sample in MRFMR, and a quantitative analysis of the material parameters by FMRFM is not straightforward when the size of the magnetic tip becomes comparable with the size of the sample.

An alternative approach to achieve large filling factors and high coupling efficiencies is to use microwave antennas whose active parts are narrow and short in conjunction with samples themselves being used as resonators. A small width of the strip-line allows for larger local microwave fields and, hence, higher microwave power-to-field conversion efficiencies \cite{Poz11boo}. A short length of the narrow, sensitive part of the strip-line allows for spatially-resolved microwave measurements. Due to the confinement of spin waves various resonant modes appear in nanoelements \cite{Hil02boo}. However, an analytical description of the magnetization dynamics in elements of arbitrary shape is barely feasible, and spin-wave eigenfrequencies can be calculated explicitly only in the few cases of simplest (e.g. circular) symmetry \cite{Sta09boo}. In particular, higher-order modes of spin-wave excitations in circular elements with radii of few hundreds nm are dipole-exchange spin waves with wavelengths $\lambda \simeq50$\,nm. Such spin waves are appreciably influenced by the exchange interaction and are, in particular, objects of current investigations in nanomagnonics \cite{Gru16nan} aiming at the operation with fast, exchange-dominated spin waves \cite{Yuh16nac,Slu19nan,Che20nac}.

\begin{figure*}[t!]
    \centering
    \includegraphics[width=0.84\linewidth]{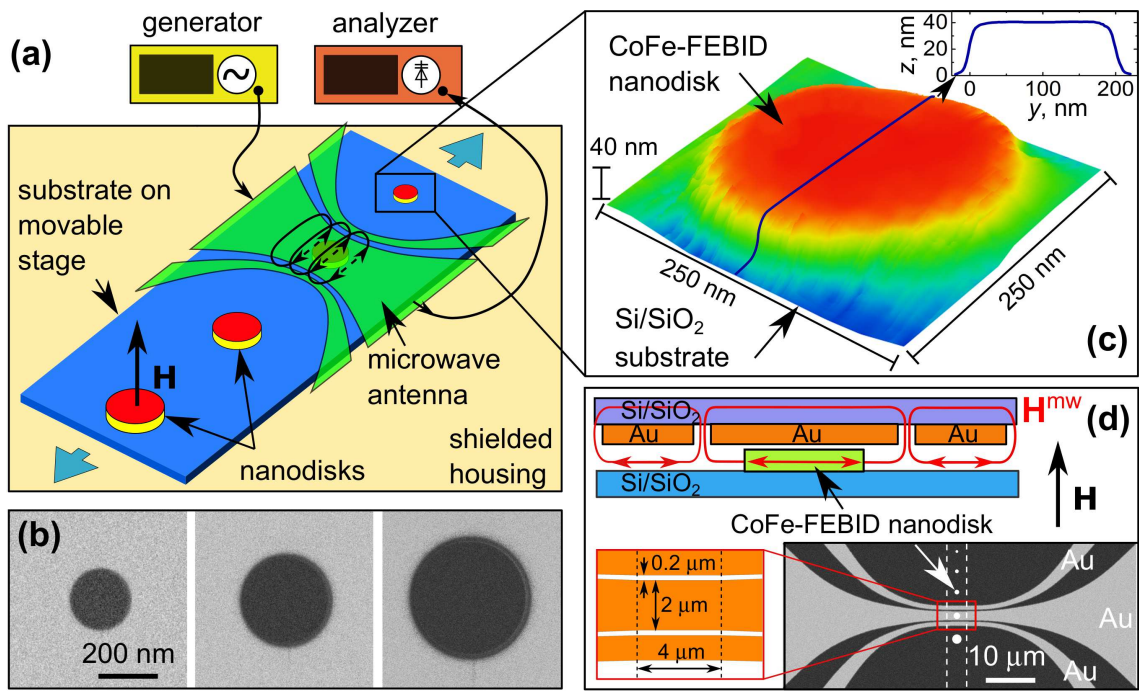}
    \caption{(a) Experimental geometry. A substrate with a series of $40$\,nm-thick circular CoFe-FEBID nanodisks is placed face-to-face to a gold microwave antenna (coplanar waveguide) for spin-wave excitation in the normal-to-disk-plane magnetic field $\mathbf{H}$. The substrate with the nanodisks is mounted onto a movable stage that allows for their precise positioning under the $4\,\mu$m-long and $2\,\mu$m-wide central conductor of the waveguide for detection of spin-wave resonances in each of the nanodisks. (b) Scanning electron microscopy images of three nanodisks with radii $R= 100, 150$, and $200$\,nm. (c) Atomic force microscopy image of the smallest nanodisk ($R = 100$\,nm) with a cross-sectional line scan in the inset. (d) Upper panel: Sketch of the microwave field $\mathbf{H}^\mathrm{mw}$ lines exciting spin waves in the adjacent nanodisk. Lower part: Scanning electron microscopy image of the inner part of the microwave antenna with the Klopfenstein tapered impedance matching sections. The dashed lines indicate the location of the nanodisk row symbolized by white solid circles.}
    \label{f1}
\end{figure*}

Here, we demonstrate spatially resolved spin-wave spectroscopy for the assessment of magnetic properties of individual circular magnetic elements with radii down to $100$\,nm. Employing a microwave antenna whose $2\times4\,\mu$m$^2$-large active part is placed over a movable substrate with well-separated circular magnetic nanoelements we detect standing spin-wave resonances from individual $40$\,nm-thick CoFe-FEBID nanodisks. These are the smallest magnetic elements successfully probed in the presence of a uniform microwave field so far. In addition to the main resonance peak, we distinguish up to $9$ higher-order spin-wave excitation modes and deduce the saturation magnetization and the exchange stiffness of individual elements with high precision by using an analytical theory. The presented approach is especially valuable for the fast on-chip characterization of individual magnetic nanoelements fabricated by FEBID whose strongest advantage is the capability of 3D nanofabrication with potential for ultrahigh density data storage and 3D magnonic networks.

The experimental geometry is shown in Figure \ref{f1}(a). The samples are circular Co-Fe nanodisks with a thickness $L$ of $40$\,nm and a radius $R$ which varies from $100$ to $1000$\,nm. Scanning electron microscopy images of the three smaller nanodisks is shown in Figure \ref{f1}(b) and an atomic force microscopy image of the nanodisk with $R = 100$\,nm is presented in Figure \ref{f1}(c). The samples were fabricated by FEBID employing HCo$_3$Fe(CO)$_{12}$ as precursor gas \cite{Por15nan}. The spin-wave excitations in the nanodisks were studied by placing the substrate with the samples in close contact face-to-face to another substrate on which a microwave antenna was prepared by e-beam lithography and covered with a $5$\,nm-thick insulating TiO$_2$ layer. The substrate with the antenna was firmly attached to a high-frequency sample holder, while an exact positioning of the individual nanodisks under the sensitive (central) part of the antenna was achieved by using a high precision piezo stage. The standing spin-wave spectroscopy measurements were done at room temperature at a fixed frequency of $f = 9.85$\,GHz with magnetic field directed perpendicular to the plane of the nanodisks. The measurements were done in a field-sweep mode, with a modulation of the magnetic field $\Delta H$ of $5$\,Oe and phase-sensitive recording of the derivative of the absorbed microwave power with respect to magnetic field.

\begin{figure*}[t!]
    \centering
    \includegraphics[width=0.9\linewidth]{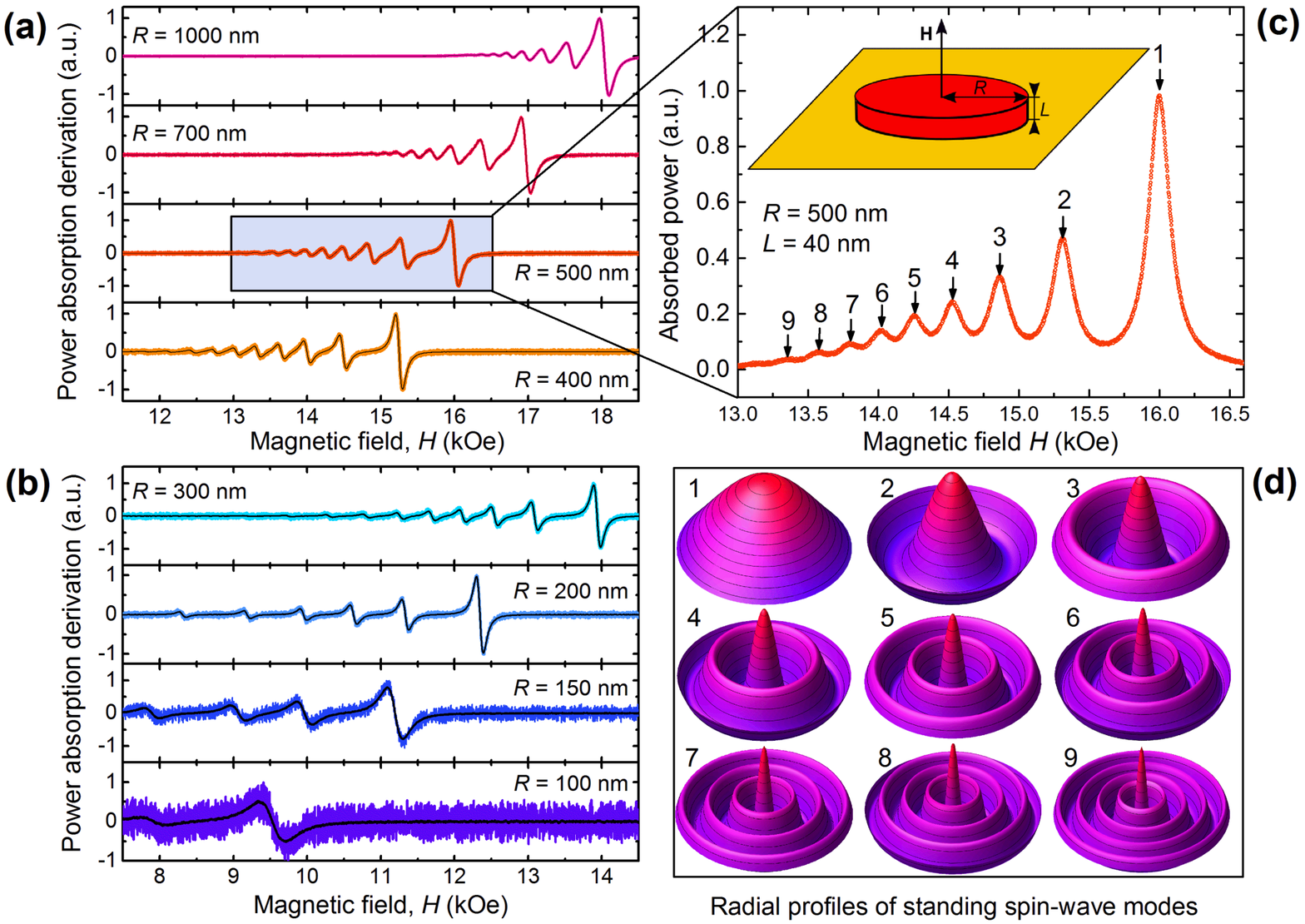}
    \caption{(a) and (b) Experimentally measured spin-wave resonance spectra at $9.85$\,GHz for a series of $40$\,nm-thick CoFe-FEBID disks with different radii, as indicated. Black lines are averages of $50$ neighboring points. (c) Field dependence of the microwave power absorption (symbols) for the disk with a radius of $500$\,nm. At least nine spin-wave resonance modes can be identified, as labeled close to the peaks. Inset: Geometry used in the analytical theory. (d) Radial profiles of the first nine standing spin-wave modes described by the zeroth-order Bessel functions.}
    \label{f2}
\end{figure*}

Figures \ref{f2}(a) and (b) present the experimentally measured spin-wave resonance spectra as a function of the out-of-plane magnetic field for the disks of different radii. For all nanodisk radii, the main resonance peak is observed at the the largest field (e.g., for the nanodisk with $R=1000$\,nm at $14$\,kOe) and it corresponds to the lowest spin-wave mode number $n = 1$, as will be detailed below. In addition to the main resonance peak there are several well-defined resonance peaks at smaller fields. The magnitude of the peaks monotonically decreases with increase of the mode number. As the disk radius decreases, the spin-wave spectra shift towards lower fields and the distance between the neighboring resonances increases.

\begin{figure*}[t!]
    \centering
    \includegraphics[width=0.98\linewidth]{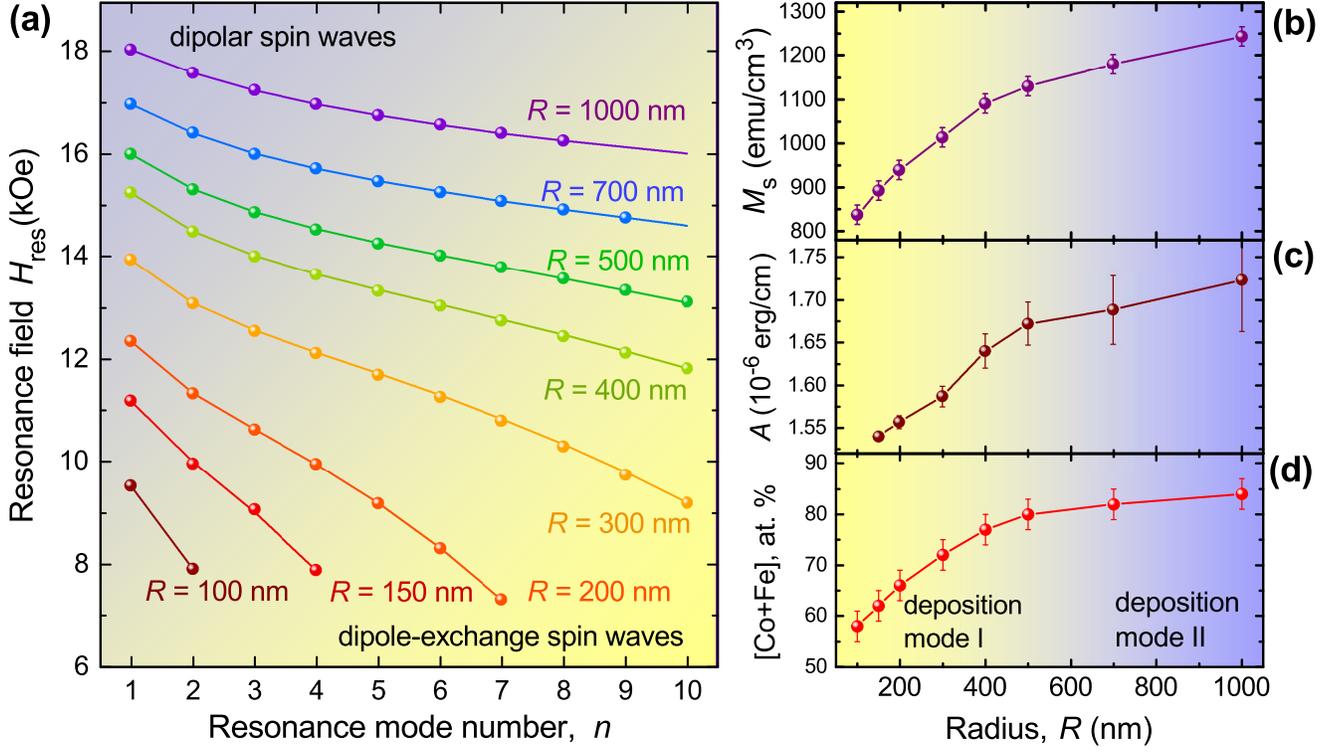}
    \caption{(a) Dependence of the resonance field $H_\mathrm{res}$ on the spin-wave mode number $n$. Symbols are experimental data. Solid lines are fits to the analytical theory with the magnetization $M_\mathrm{s}$ and the exchanged constant $A$ varied as fitting parameters, as reported in panels (b) and (c), and the gyromagnetic ratio $\gamma/2\pi = 3.05$\,MHz/Oe. The background color gradient designates a crossover from the regime of dipolar spin waves at larger $R$ and smaller $n$ to the regime of dipole-exchange spin waves at smaller $R$ and larger $n$. (d) Metal content in the nanodisks as inferred from energy-dispersive X-ray spectroscopy. The lines in (b), (c) and (d) are guides for the eye. The background color in (b), (c), and (d) indicates a crossover from the deposition of samples in a depleted-precursor mode (deposition mode I, yellow) to an almost precursor depletion-free mode (deposition mode II, blue) as the disk size increases.}
    \label{f3}
\end{figure*}

Importantly, it was possible to detect the signal even for the smallest disk with a radius of only $100$\,nm. In previous studies, arrays consisting of $10^6$--$10^7$ identical disks were investigated using conventional FMR spectroscopy to reach the required sensitivity. In our experiment, for the smallest nanodisk the signal-to-noise ratio (SNR) is about $2$. Nevertheless, two spin-wave peaks are observable at the positions predicted by the analytical theory to be introduced in what follows. For a larger nanodisk with $R = 150$\,nm, the SNR increases to about $5$ and four spin-wave peaks can be clearly distinguished. The SNR continues to noticeably increase as the disk radius $R$ increases, and more and more standing spin-wave modes become visible. For instance, the largest number of modes ($9$ modes) can be identified for the disk with $R = 500$\,nm, as illustrated in Figure \ref{f2}(c). From the measured spin-wave spectra, for each mode $n$ and disk radius $R$, the resonance fields $H_\mathrm{res}$ were deduced at fields at which the power absorption derivative turns into zero, which is equivalent to the definition of $H_\mathrm{res}$ from at the absorbed power maxima in Figure \ref{f2}(c). The deduced dependences of the spin-wave resonance fields $H_\mathrm{res}$ on the mode number $n$ for different nanodisk radii are presented in Figure \ref{f3}(a).

Standing spin waves appear in nanoelements because of the quantization of the radial spin-wave vector component due to the elements' finite lateral size \cite{Kak04apl}. To describe analytically the field values at which the spin-wave resonance appear, we consider azimuthally symmetric spin waves in a thin cylindrical ferromagnetic disk of thickness $L$ and radius $R$ saturated in the out-of-plane direction by the biasing magnetic field $H$, see the inset in Figure \ref{f2}(c). In the case of out-of-plane magnetized circular disks the excited spin-wave eigenmodes can be described by Bessel functions of the zeroth order because of the axial symmetry of the samples \cite{Kak04apl}. For details on the analytical calculations we refer to the Methods section. The radial profiles of the first nine spin-wave modes are illustrated in Figure \ref{f2}(d).

Fitting of the experimental dependences $H_\mathrm{res}(n)$ in Figure \ref{f3}(a) with the use of the theoretical model allows for the deduction of the saturation magnetization $M_\mathrm{s}$ and the exchange stiffness $A$ for all individual nanodisks (see Figures \ref{f3}(b) and \ref{f3}(c)). In Figure \ref{f3}(a), solid lines are fits to the analytical theory with the magnetization $M_\mathrm{s}$ and the exchanged constant $A$ varied as fitting parameters, as reported in Figures \ref{f3}(b) and (c), and the gyromagnetic ratio $\gamma/2\pi = 3.05$\,MHz/Oe \cite{Tok15prl}. Specifically, the value of $A$ affects the decrease rate of $H_\mathrm{res}(n)$ in Figure \ref{f3}(a) (the larger $A$, the faster the decrease of $H_\mathrm{res}(n)$) while the location of the main resonance peak is determined by a combination of $M_\mathrm{s}$ and $A$ (the higher $M_\mathrm{s}$, the larger $H_\mathrm{res}(n)$). The deduced $M_\mathrm{s}$ and $A$ values suggest that in contrast to samples prepared by electron-beam lithography, $M_\mathrm{s}$ of the nanoelements fabricated by FEBID significantly depends on the elements' lateral dimensions, and even for the largest nanodisk with $R = 1000$\,nm, the deduced $M_\mathrm{s}$ value is smaller than $M_\mathrm{s}$ of bulk Co.

\enlargethispage{1\baselineskip}
We believe that the decrease of $M_\mathrm{s}$ with decrease of the disk radius is a consequence of the employed writing strategy in the FEBID process. This underlines the importance of the characterization described here for as-fabricated FEBID samples. Namely, in the FEBID process, each nanodisk was defined as a circular polygon in which the number of points for the beam to dwell is proportional to the square of the nanodisk radius. The writing of the $40$\,nm-thick nanodisks required a few thousands passes of the electron beam. Accordingly, given that the point-to-point distance (pitch) for all nanodisks was kept constant, a complete rastering of the beam over the smaller disks occurred faster as compared to the larger disks. Because of the finite time needed for the precursor gas to replenish in the vicinity of the sample, the smaller disks were written in the depleted-precursor regime, which results in an increase of the carbon and oxygen content at the expense of cobalt and iron. A crossover from the deposition of samples in the depleted-precursor mode (deposition mode I) to an almost precursor depletion-free mode (deposition mode II) with increase of the disk size is illustrated by the yellow-to-blue background color gradient in Figure \ref{f3}(b)-(d). We note that a decrease of $M_\mathrm{s}$ with decrease of the size of Co-FEBID nanospheres grown on cantilever tips was also attributed to a decrease of the metal content in the smaller spheres \cite{San17bjn}.

\vspace{5mm}
Specifically, for the samples investigated here, the [Co+Fe] content in the nanodisks has been found to decrease from about $83$\,at.\% for the largest disk to about $58$\,at.\% for the smallest disk while the [C+O] content increased from about $17$\,at.\% to about $42$\,at.\%, see Figure \ref{f3}(d). Remarkably, the deduced $M_\mathrm{s}$ values in Figure \ref{f3}(b) correlate rather well with the [Co+Fe] content in Figure \ref{f3}(d) which suggests that the reduction of $M_\mathrm{s}$ in the samples can be attributed to the reduction of the content of the magnetic Co and Fe in the sample volume. By employing a different writing strategy in which the electron beam was ``parked'' for $10$\,ms outside of the structure after each pass to allow the precursor to replenish, $M_\mathrm{s}$ values close to $1350$\,emu/cm$^3$ have been deduced for a reference set of samples regardless of the nanodisk size. Meanwhile, the deduced values of $A(R)$, despite following qualitatively the behavior of $M_\mathrm{s}(R)$, are decreasing by only about $10\%$ with decrease of the disk radius. In Figure \ref{f3}(c), the deduced $A$ value for the nanodisk with $R=100$\,nm is not shown because only two modes were observed for this sample, that results in a relatively large error in the determination of the exchange stiffness.

\begin{figure*}[t!]
    \centering
    \includegraphics[width=0.98\linewidth]{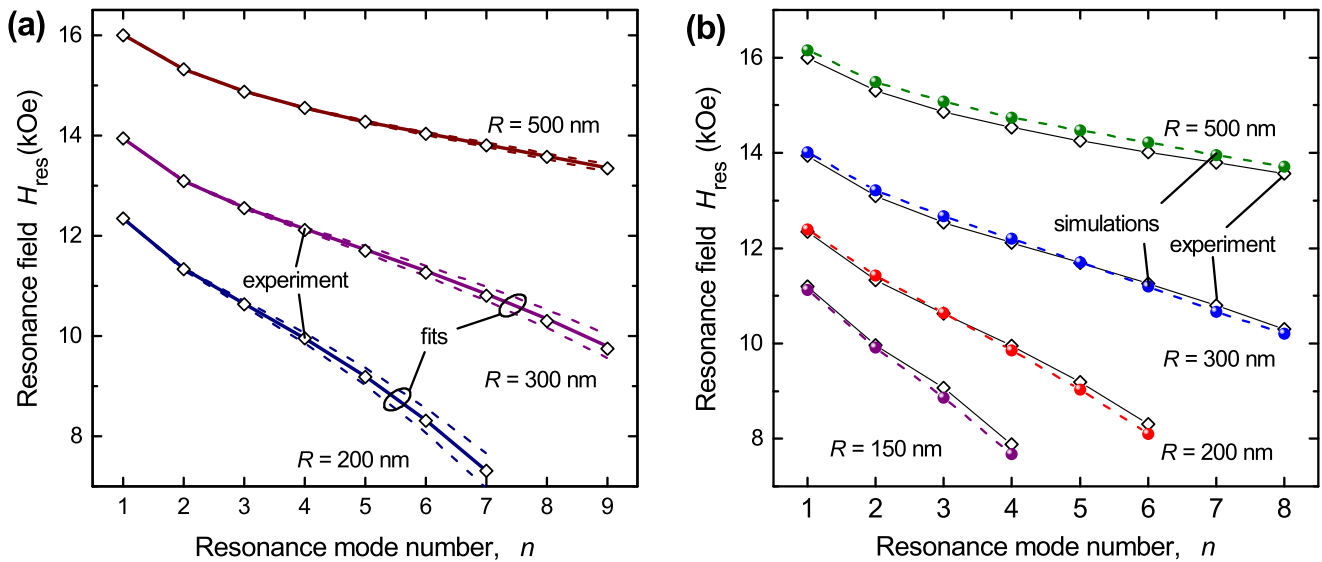}
    \caption{(a)~Fits of the experimental spin wave resonance spectra for the disks with different radii using the same values of $M_\mathrm{s}$ and $A$ as in Figure \ref{f3}, but also for different $A$ values. Solid lines correspond to the best-fit $A$ values in Figure \ref{f3}(c) while dashed lines pertain to $A$ values that are smaller or larger than the best-fit ones by $10\%$.
    (b) Comparison of the experimental and simulated resonance fields $H_\mathrm{res}$ for a series of samples and resonance modes. Micromagnetic simulations were performed using MuMax3, with the saturation magnetization $M_\mathrm{s}$ and the exchange stiffness $A$ as parameters deduced from the fits of the experimental data to the analytical theory in Figure \ref{f3}.}
    \label{f4}
\end{figure*}

Figure \ref{f4}(a) illustrates the accuracy of the fits of the experimental dependences $H_\mathrm{res}(n)$ to the analytical theory using the same values of $M_\mathrm{s}$ and $A$ as in Figures \ref{f3}(b) and (c) in comparison with the curves for different $A$ values. Solid lines in Figure \ref{f4}(a) correspond to the best-fit $A$ values in Figure \ref{f3}(c) while dashed lines pertain to $A$ values that are smaller or larger than the best-fit ones by $10\%$. As follows from the fits, the accuracy of the deduction of $A$ strongly increases with reduction of the nanodisk radius. This is a consequence of the increasing role of the exchange interaction affecting the spin-wave dynamics in smaller nanodisks. In addition, the exchange interaction more strongly influences higher-order spin-wave modes. This can be understood as a consequence of the spin-wave wavelength reduction with increase of $n$ and the associated crossover from the regime of dipolar spin waves at larger $R$ and smaller $n$ (with spin-wave wavelengths $\lambda \backsimeq 1\,\mu$m) to the regime of dipole-exchange spin waves at smaller $R$ and larger $n$ (with $\lambda \lesssim 100$\,nm). This crossover is indicated by the background color gradient in Figure \ref{f3}(a). Specifically, in our measurements the shortest wavelengths were detected for the disks with $R=200$\,nm and $300$\,nm. The mode numbers $n=7$ and $n=9, 10$ detected for $R=200$\,nm and $R=300$\,nm, respectively, correspond to effective spin-wave wavelengths of $\lambda \simeq R/n \approx 30$\,nm whose properties are determined by both, the dipolar and exchange interactions.

The reliability of the magnetic parameters deduced from the analytical theory was additionally examined by means of micromagnetic simulations. The simulations were done using MuMax3 \cite{Van14aip} for four nanodisks with radii of $150$, $200$, $300$, and $500$\,nm. The simulation results are presented in Figure \ref{f4}(b), where the simulated $H_\mathrm{res}$ values as a function of the spin-wave mode number $n$ fit well the experimentally measured ones.
The disagreement between the simulated and the experimentally measured $H_\mathrm{res}$ values does not exceed $3\%$. This means that the used analytical theory is reliable in the whole range of aspect ratios $\beta = L/R$ of up to $0.4$ and it can be applied for the deduction of the saturation magnetization and the exchange stiffness for nanomagnets of cylindrical shape. Once $M_\mathrm{s}$ and $A$ are deduced, they can be further used as input parameters for simulations (see Supporting Information Note 1, Figure S1 and Movie 1) which allow one to predict the response of the nanodisks at essentially higher frequencies than $9.85$\,GHz used just as an exemplary FMR frequency in our experiment.

Turning to the general importance of the obtained results, it is worth noting that a particular advantage of the employed standing spin-wave resonance technique in comparison with conventional magnetometry is that it is not necessary to know the mass of the sample for the deduction of its $M_\mathrm{s}$. The knowledge of the geometrical dimensions of the sample is sufficient. This is, in particular, important in the case of samples fabricated by FEBID since the density of the deposits may vary depending on the deposition parameters and post-growth treatments such as annealing and irradiation with ions and electrons \cite{Dob15bjn}. Also, such an important parameter as the exchange stiffness can be deduced from fits of the experimental data to the analytical theory. In return, the deduced saturation magnetization and the exchange stiffness can be used as input parameters for simulations of more complex 3D nanoarchitectures \cite{Hut18mee,Kel18nsr,Win19jap,Sko20nal,San20arx}. This is especially fortunate, because various complex geometries with lateral feature sizes down to $10$\,nm can be fabricated by FEBID, but no analytical theory is available for their description despite the few cases of simplest symmetries. In this way, due to the available analytical description allowing for precise deduction of $M_\mathrm{s}$ and $A$, circular nanodisks can be used as standard reference samples for the verification of micromagnetic simulations which, in turn, allow for modeling the magnetization dynamics in complex-shaped direct-write nanostructures for 3D nanomagnetism and spintronics. In addition, the capability of FEBID to deposit magnetic layers with pre-defined properties onto conventional microwave circuits makes it a valuable fabrication tool for nanomagnonics requiring microwave-to-magnon transducers for excitation and detection of phase-coherent exchange magnons with $\lambda$ below $50$\,nm \cite{Che20nac}.

In summary, we have demonstrated scanning spin-wave spectroscopy of individual circular magnetic elements with radii down to $100$\,nm and total volumes down to $10^{-3}\,\mu$m$^3$. These are the smallest magnetic elements successfully probed in the presence of a uniform microwave field so far. We have illustrated the use of standing spin waves as magneto-dynamical probes of nanodisks fabricated by FEBID and exhibiting a decrease of $M_\mathrm{s}$ with decrease of the disk radius because of a reduction of the metal content with decrease of the elements' sizes. Up to 9 higher-order modes of spin-wave resonances have been observed and effective spin-wave wavelengths down to $30$\,nm have been detected. The two key ingredients of the described approach are the microwave antenna and the circular symmetry of the nanodisks allowing the samples themselves to act as nanoresonators. Furthermore, the circular symmetry of the nanodisks has allowed for the deduction of the saturation magnetization and the exchange stiffness of the material with high precision using an analytical theory. The deduced values have been used as input parameters for micromagnetic simulations which allowed us to reproduce the experimentally measured resonance fields with an accuracy better than $3\,\%$ and, hence, can be used for modeling the magnetodynamic response of complex-shaped magnetic nanoelements for which no analytical description is available. In all, the developed methodology is especially valuable for the characterization of building blocks of direct-write nano-architectures for 3D nanomagnetism and magnonics.

\begin{figure*}[t!]
    \includegraphics[width=0.9\linewidth]{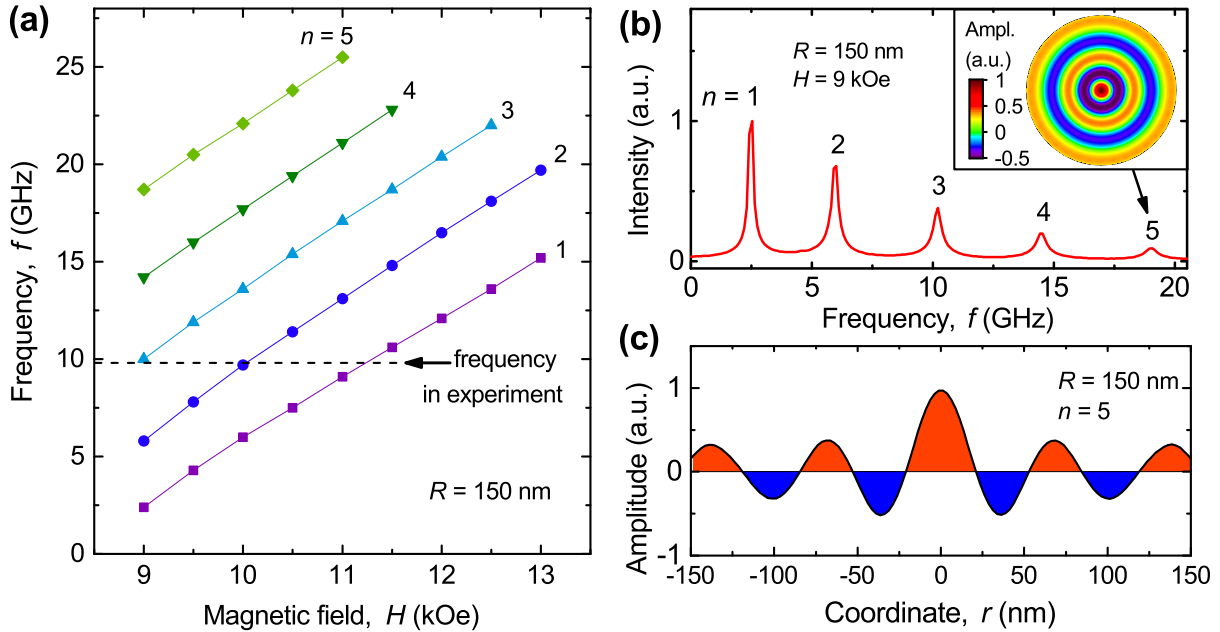}
    \caption{(a) Simulated dependences of the resonance frequencies on the magnetic field for the first five resonance modes for the disk with $R = 150$\,nm.
    (b) Simulated spin-wave spectrum for the nanodisk with $R = 150$\,nm at $9$\,kOe.
    Inset: Spatial dependence of the standing spin-wave amplitude for the fifth mode ($n = 5$) in the nanodisk.
    (c) Radial profile of the spin-wave amplitude for $n=5$.}
\end{figure*}

This work was supported by the European Commission in the framework of the Horizon 2020 Marie Sklodowska-Curie program--Research and Innovation Staff Exchange (MSCA-RISE) under Grant Agreement No. 644348 (MagIC).
Further, support by the European Cooperation in Science and Technology via COST Action CA16218 (NANOCOHYBRI) is acknowledged.
The Portuguese team acknowledges the Network of Extreme Conditions Laboratories-NECL and Portuguese Foundation of Science and Technology (FCT) support through Project Nos. NORTE-01-0145-FEDER-022096, POCI-0145-FEDER-030085 (NOVAMAG), and EXPL/IF/00541/2015.
D.N. acknowledges the Spanish Ministry for Science, Innovation and Universities, for funding through the ``Ramon y Cajal'' program RYC-2017-22820.
A.V.C. acknowledges support within the ERC Starting Grant no. 678309 MagnonCircuits.
M.K. and P.G. acknowledge support of National Science Center Poland under the project  UMO-2018/30/Q/ST3/00416.
K.Y.G. acknowledges support by IKERBASQUE (the Basque Foundation for Science) and by the Spanish Ministry of Economy and Competitiveness under the project FIS2016-78591-C3-3-R.
Support through the Frankfurt Center of Electron Microscopy (FCEM) is gratefully acknowledged.

Supplementary Figure 5 presents simulation results for the nanodisk with $R = 150$\,nm. The resonance frequencies $f(H)$ increase almost linearly with increasing magnetic field, see Figure 5(a). The frequency separation between the neighboring resonances amounts to about $4.5$\,GHz, as is illustrated in Figure 5(b) for $H=9$\,kOe which is very close to the resonance field for the third mode in the experiment. We have also checked that the micromagnetic simulations nicely reproduce the profiles of the spin-wave amplitude described by the zeroth-order Bessel functions in the analytical theory. For instance, if one takes a look at the fifth resonance mode, from the spatial profile of the spin-wave amplitude in Figure 5(c) follows that the involved spin wave length $\lambda \simeq 65$\,nm and such spin waves are in the dipole-exchange regime. At yet higher frequencies (and larger magnetic fields), higher excitation modes in the smallest nanodisks are represented by exchange-dominated spin waves.

\textbf{Author Contributions}.
O.V.D. and G.N.K. conceived and supervised the project.
O.V.D., G.N.K. and M.H. designed the samples.
O.V.D., R.S., S.A.B. and G.N.K. fabricated the samples and performed the measurements.
S.A.B., N.R.V., D.N., P.G. and M.K. performed micromagnetic simulations and compared their results with the experimental data.
S.A.B. and N.R.V. fitted the experimental data to the analytical theory.
K.Y.G. provided theoretical support.
O.V.D., M.K., M.H., A.V.C., K.Y.G. and G.N.K. discussed the interpretation and the relevance of the results.
O.V.D. wrote the first version of the manuscript with contributions from all authors.

\section*{Methods}
\textbf{Fabrication of nanomagnets}. The samples were fabricated by focused electron beam induced deposition (FEBID) in a high-resolution dual-beam scanning electron microscope (SEM: FEI Nova NanoLab 600). FEBID represents the process by which a metal-organic precursor gas, in this case HCo$_3$Fe(CO)$_{12}$, adsorbed on a substrate surface, is dissociated in the focus of the electron beam into a permanent deposit and volatile components \cite{Hut18mee}. Deposition occurs only where the beam focus dwells for a longer time (typically $500$\,ns up to several ms) on a given point on the substrate surface in accordance with a pre-defined pattern. In addition to the high resolution, which is down to $10$\,nm laterally and $1$\,nm vertically for the precursor gas used here, FEBID is capable of creating binary and ternary alloy systems and complex 3D nano-architectures \cite{Kel18nsr,Win19jap,Sko20nal,San20arx}. The precursor was injected into the SEM through a capillary with an inner diameter of $0.5$\,mm. The distance capillary-surface was about $100\,\mu$m and the tilting angle of the injector was $50^\circ$. The crucible temperature of the gas injection system was $65^\circ$C. The disks were grown in the high-resolution deposition mode with $5$\,keV beam energy, $1.6$\,nA beam current, $20$\,nm pitch, and $1\,\mu$s dwell time. The beam was scanning in a serpentine fashion. On both sides of the nanodisk row, a few $10\times10\,\mu$m$^2$ insulating TiO$_2$-FEBID layers were deposited as spacers for mechanical protection of the nanodisks.

\textbf{Compositional analysis}. The elemental composition of the largest disk with $R = 1000$\,nm is about 83\,at.\% [Co+Fe], 10\,at.\% [O] and 7\,at.\% [C], as inferred from energy-dispersive X-ray (EDX) spectroscopy. The EDX test area was $100\times100$\,nm$^2$, and the EDX parameters were $5$\,kV and $1.6$\,nA. Here, the beam energy determines the effective thickness of the layer being analyzed, which is approximately $45$\,nm, thus a minor part of the counts for oxygen should be attributed to the topmost SiO$_2$ layer on top of the Si substrate. The penetration of the electrons into the film was calculated by the simulation program CASINO. This corresponds to approximately $90\%$ of the electron beam energy being dissipated in the nanodisks. The material composition was calculated taking ZAF (atomic number, absorbtion, and fluorescence) and background corrections into account. The software used to analyze the material composition of the samples was EDAX's Genesis Spectrum v.5.11. The statistical error in elemental composition is $5\%$, as depicted by the error bars in Figure \ref{f3}(d).

\textbf{Microstructural characterization}. A previous microstructural investigation of the Co-Fe deposits by transmission electron microscopy revealed that they consist of a dominating bcc Co$_3$Fe phase mixed with a minor amount of FeCo$_2$O$_4$ spinel oxide phase with nanograins of about $5$\,nm diameter \cite{Por15nan}. The thickness of all nanodisks was $L = 40\pm1$\,nm while their radii $R$ were $100$, $150$, $200$, $300$, $400$, $500$, $700$, and $1000$\,nm. The deposited nanodisks were arranged in a row, with a nanodisk center-to-center distance of $5\,\mu$m. This distances was selected to avoid the dipolar coupling between the nanodisks and to avoid the detection of resonance signals from nanodisks which are not immediately under the microwave antenna. For high-resolution characterization of the fabricated nanodisks, atomic force microscopy (AFM) under ambient conditions in the noncontact, dynamic force mode was used. The cantilever tip was shaped like a polygon-based pyramid, with a tip radius of less than $7$\,nm (Nanosensors PPP-NCLR), so that convolution effects due to the finite tip radius can be neglected. A root-mean-square surface roughness of less than $0.2$\,nm over a scan field encompassing the entire nanodisk surface area was inferred from the AFM inspection. Around all nanodisks, within about a $20$\,nm-wide area, an up to $3\,$nm-thick co-deposit layer has been revealed. This layer is non-magnetic and it appears as a consequence of the finite emission cone of secondary electrons in the FEBID process.

\textbf{Microwave antenna}. Ferromagnetic resonance measurements were performed in the flip-chip geometry in which the substrate with the nanomagnets was placed face-to-face to the substrate containing a microwave antenna (coplanar waveguide) prepared by e-beam lithography from an Au film. The $55$\,nm-thick Au film was prepared by dc magnetron sputtering onto a Si/SiO$_2$ substrate with a pre-sputtered $5$\,nm-thick Cr buffer layer. The thickness of the SiO$_2$ layer was $200$\,nm. In the sputtering process, the substrate temperature was $T = 22^\circ$C, the growth rates were $0.055$\,nm/s and $0.25$\,nm/s, and the Ar pressures were $2\times10^{-3}$\,mbar and $7\times10^{-3}$\,mbar for the Cr an Au layers, respectively. Spin-wave resonances were excited and detected by the sensitive central part of the waveguide, as illustrated in Figure \ref{f1}(d). The width of the signal and ground conductors of the coplanar waveguide was $w = 2\,\mu$m and $1\,\mu$m, respectively, while the gap between them amounted to $g = 200$\,nm, resulting in a microwave impedance $Z$ of about $90\,\Omega$ at $9.85$\,GHz. The $2\,\mu$m-wide central conductor of the coplanar waveguide ensures that the nanodisks with $R<1000$\,nm are excited in the presence of a uniform microwave field. The impedance of the antenna was matched to the $Z_0 = 50\,\Omega$ impedance of the transmission line by two Klopfenstein tapers \cite{Poz11boo} designed for keeping the magnitude of the microwave power reflection coefficient below $0.05$ at frequencies $f > 1\,$GHz. The central part of the antenna was covered with a $5$\,nm-thick TiO$_2$ FEBID layer for electrical insulation of the antenna from the samples. The fine positioning of individual nanodisks under the sensitive part of the antenna was achieved by using a Tritor 38 3D translational stage allowing for a spatial positioning accuracy of $1$\,nm.

\textbf{Ferromagnetic resonance measurements}. The spin-wave spectroscopy measurements were done at a fixed frequency of $f = 9.85$\,GHz with magnetic field directed perpendicular to the plane of the nanodisks. The misalignment error of the field angle was less than $0.1^\circ$. The high-frequency ac stimulus was provided by a continuous wave microwave generator and the transmitted microwave signal was detected by a signal analyzer. The measurements were done in a field-sweep mode, with a modulation of the magnetic field $\Delta H$ of $5$\,Oe, field modulation frequency of $15$\,Hz, and a phase-sensitive recording of the derivative of the absorbed microwave power. This allowed for an increase of the measurement sensitivity by up to about two orders of magnitude in comparison with the standard detection regime in which a fixed external field $H$ is applied and the frequency is swept without modulation. The coplanar waveguide antenna allows for measurements in a broad range of microwave frequencies. However, in this work we performed measurements at a fixed frequency as this was sufficient for the performed analysis. For a microwave power of $10$\,mW the amplitude of the magnetic field produced at a distance of $25$\,nm from the antenna is estimated as $40$\,Oe. Before measurements, all samples were saturated in an out-of-plane field of $20$\,kOe.

\textbf{Analytical calculations}. Standing spin waves appear in nanoelements because of the quantization of the radial spin-wave vector component due to the elements' finite lateral size \cite{Kak04apl}. The spin-wave dispersion relation for an infinite film has the form $\omega^2 = \omega_H(\omega_H + \omega_M F(kL))$ \cite{Kal86jpc}, where $\omega_H = \gamma(H + 2Ak^2/M_\mathrm{s})$, $\omega_M = \gamma 4\pi M_\mathrm{s}$, $\gamma$ is the gyromagnetic ratio, $A$ is the exchange stiffness, $M_\mathrm{s}$ is the saturation magnetization of the material, and $k$ is the in-plane wave vector component. The function $F(x)$ describes the dynamical dipolar interaction and can be approximated as $F(x) = 1- (1-\exp(-x))/x$. In the case of out-of-plane magnetized circular disks the excited spin-wave eigenmodes can be described by Bessel functions of the zeroth order because of the axial symmetry of the samples. The dynamic in-plane magnetization components are $m_x, m_y \propto J_0 (k\rho)$, where $\rho$ is the radial coordinate. Assuming that the static disk magnetization does not depend on the thickness coordinate $z$, one can use a spin-wave dispersion equation similar to the one used for infinite films \cite{Kak04apl}
\begin{equation}
\label{e1}
    \omega^2_n = \omega_H(\omega_H + \omega_M F(\kappa_n \beta)),
\end{equation}
where $\omega_n$ is the frequency of the spin-wave mode with the quantized radial wave vector $k_n = \kappa_n/R$ and magnetic field $H$ is substituted by the effective magnetic field $H + H_{nn}$, where $H_{nn}$ are the diagonal matrix elements of the static dipolar field using a complete set of eigenfunctions as defined below, and $\beta = L/R$ is the thickness-to-radius aspect ratio of the nanodisk. The disk aspect ratio $\beta = L/R$ is small and the dipolar pinning of the dynamical magnetization at the disk edges is strong \cite{Gus05prb,Kak12prb}. This results in the simple relation $\kappa_n = \alpha_n$, where $\alpha_n$ is the $n$-th root of the equation $J_0(\kappa) = 0$, $n=1,2,3,...$. The normalized eigenfunctions
$\phi_n(\rho) = \displaystyle\frac{1}{\sqrt{N_n}}{J_0}({\kappa_n} \rho)$ and
$N_n = \frac{1}{2}[J_0^2(\kappa_{n}) + J_1^2 (\kappa_n)]$ lead to the following matrix elements $H_{nn}$ of the dipolar field
\begin{equation}
    \label{e2}
     \frac{H_{nn}}{4\pi M_\mathrm{s}} = \int_0^\infty dxf(\beta x) J_1(x)
     \int_0^1 d\rho\,\rho \phi^2_n(\rho) J_0(\rho x)-1.
\end{equation}

\textbf{Micromagnetic simulations}. The micromagnetic simulations were performed by using the MuMax3 solver \cite{Van14aip}. The $M_\mathrm{s}$ and $A$ values were extracted from the best fits of the experimental data to the described analytical theory. The values reported in Figures \ref{f3}(c) and (d) have been used as the input parameters for the simulations. For all nanodisk radii the cell size was $2.5\times2.5\times2.5$\,nm$^3$ and the damping parameter was $\alpha = 0.01$. The simulations were performed in two stages, a static one and a dynamic one. At the first stage, an equilibrium magnetic configuration of the system was reached by relaxing a random magnetic configuration for each value of the out-of-plane bias field $H$ from a given set. Subsequently, magnetization precession was excited by applying a small spatially uniform in-plane microwave field $h\propto \mathrm{sinc}(2 \pi f_\mathrm{cut}(t-t_0))$, where $f_\mathrm{cut} = 40$\,GHz  is the cut-off frequency and $t_0=0.5$\,ns is the time delay after which the sinc function reaches its maximum. The dynamic response of the system has been recorded with the sampling interval equal to $1.25$\,ns. Finally, a fast Fourier transform was used to extract the normalized frequency spectra and the magnetization profiles at the resonance frequencies of the standing spin waves.

%

\end{document}